# The perverse incentive for default swap instruments that are derivatives – solving the jackpot problem with a clawback lien for equity default swaps.

**Brian P. Hanley**
**Brian.Hanley@ieee.org**

## Abstract

When an equity swap contract is also a derivative a serious problem arises because a derivative must be fulfilled immediately. This feature of derivatives prevents claims processing procedures that screen out ineligible claims in the ordinary insurance industry. This, in turn, creates a perverse incentive for holders of such notes to commit fraudulent acts that result in payment. This problem first surfaced with credit default swap (CDS) contracts, which are part of a class of swap instruments I term default swaps.

Without an address to this problem, within 90% of venture capital firms, a venture-bank makes less money the better their investments do, in a continuous function. Thus, for the majority of venture firms the highest rate of return they can make is to cause total loss intentionally.

Here, a strategy for removing this perverse incentive is defined, consisting of a lien on the firm that is the beneficiary of the insurance. This clawback lien may be for as much as 100% of the insurance payment, payable at a later date. This is the final component for implementing the Equity Default Clawback Swap (EDCS) derivative instrument. Removing the perverse incentive minimizes reasons for underwriters to deny EDCS coverage to new and lower performing venture capital firms, which are 90% of all venture capital firms.

## 1 Introduction

There is a long-standing problem in the insurance industry with bad-faith clients. Because of this problem, a suicide will not result in payout of the claim, nor will an arsonist be paid for damage to property. Similarly, a diamond merchant will not be paid for diamonds fenced on the black market. This is dealt with in classical insurance using riders denying claims where the client causes their own insured harm intentionally. The denial of claims requires a claim processing department and investigators that examine each claim and decide whether a claim will be paid or not, which can take some time and may entail litigation.

There is, however, a serious problem that arises when such an instrument is a derivative. Swap instruments like the CDS inhabit an area where they are not classified as insurance, but for the banks that buy them they are accepted as insurance. They do not, however, function similarly for the sellers. Since they are derivatives, the law provides for immediate execution, without recourse to the courts. The USA's 1982 safe-harbor bankruptcy code provisions were the first to grant derivative holders the right to immediate foreclosure on underlying assets (Gilbane 2010). The history of such safe-harbor provisions for forward, commodity, and security contracts, repurchase and swap agreements is founded on necessity.

Consider, for instance, what would happen if the holder of an orange juice future contract could be held off by court filing. Let us say that the price of orange juice rose, which gives the futures call option a profit. Then let us say that the owner of the orange juice refuses to honor the call option, sells the orange juice to a food company and pockets the difference. This forces the holder of the call option to go to file in court, at his expense, and wait for the case to be decided. A smart litigator can draw things out for a long time. This would end meaningful trade in futures contracts, thereby destroying the market for them. Thus, the law does not allow this.

Public law 109-8, also known as BACPA, passed in 2005 (Grassley 2005), ensures new derivative instruments are treated in the same way.

This derivative-classed insurance instrument jackpot problem was first seen in the run-up to the 2008 banking crisis (Hanley, 2012). In that crisis, AIG was the underwriter, and various major banks in the USA were the clients purchasing credit default swap (CDS) contracts for home loans. Citigroup, for instance, was accused of lying to investors who bought securities composed of loans. Citigroup had bought CDS's against the loans (Rakoff 2011; Wyatt 2011), and retained the CDS's after selling off the loans. When the loans went into default, Citigroup collected on the CDS contracts. I have termed such insurance on loans that are derivatives, an equity default clawback swap (EDCS)

It doesn't take much thought to see that it is quite profitable to make a poor quality loan, buy a swap instrument on that loan, sell off the loan, then collect on the swap instrument. Doing so returns capital in the initial sale, then doubles capital when the loan defaults. If half the loans in a bundled security default within 12 months,, then the overall return is 1.5, or 50% for that year. AIG got caught by that perverse incentive (Hanley, 2012).

## 2 Venture-banking overview

Venture-banking is a new concept I have defined that uses underwriting of an equity default clawback swap (EDCS) to insure investments that are made through bank loans (Hanley, 2017). The use of the EDCS allows the bank to book the insured value of the EDCS back into Tier 1 or Tier 2 reserves. I show that an underwriter can operate quite profitably on broad venture capital portfolios within a range of conventional returns that are achieved by real world portfolios. I also show that venture banking dramatically increases rates of return. The primary data source I used came from Ewing Marion Kauffman Foundation's 20 years of venture capital experience in banking, as shown in figure 1 (Mulcahy, 2012).

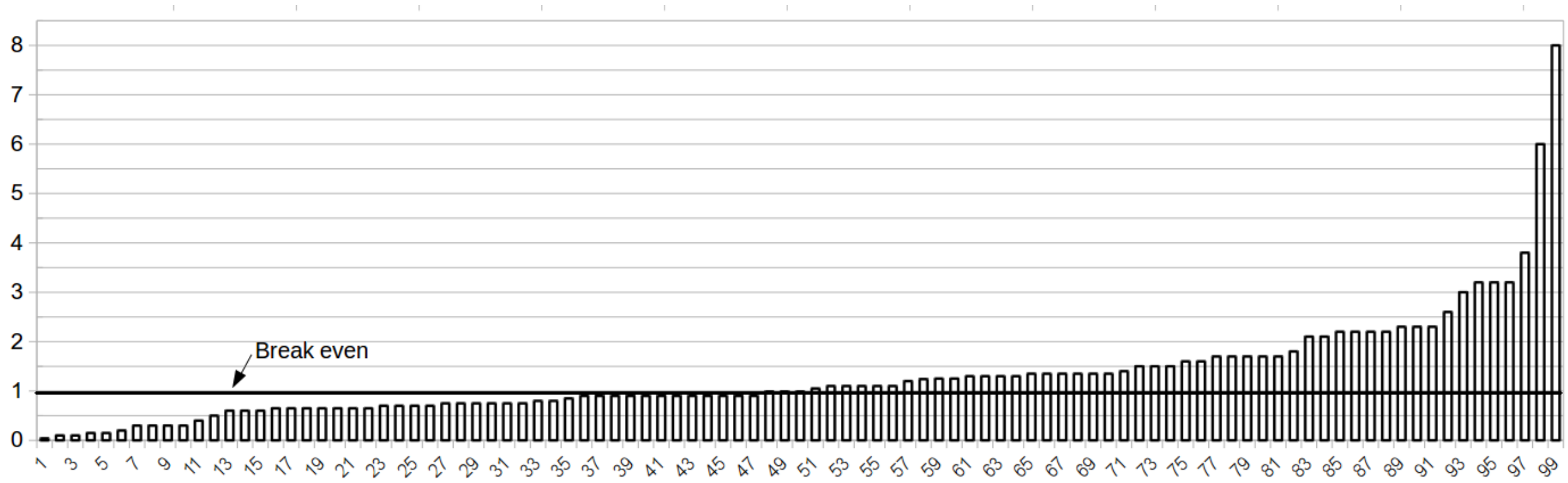

*Figure 1: Kauffman Foundation, 20 years of venture capital returns on 99 funds.*

The EDCS instrument for insuring venture-capital administered as bank loans enables use of the EDCS to take advantage of the multiplier I discovered analyzing the 2008 banking crisis (Hanley, 2012). Unlike the classical banking multiplier[1], this multiplier is only limited by the regulatory limits on Tier 1 and Tier 2 capital composition. In the venture-banking design, the EDCS is a default swap instrument to replenish bank capital. Like the CDS, an EDCS is effectively insurance for the buyer and accepted as such by regulators. Insuring the loan allows the EDCS to be accounted for as Tier 1 or Tier 2 capital. This allows each dollar of original capital in the venture-bank to be multiplied by up to 47 times, without needing to access the Federal Reserve to replenish reserves. The actual multiple of the original capital (MOC) can vary from 1 to 47. In my modeling I chose a low of 30 and a high of 43 as the normative MOC values. An MOC of 43 still provides significant headroom in a crisis. For more conservative venture-bank managers, an MOC of 30 still provides excellent returns while maintaining nearly 50% potential reserve capital.

The core of the venture-banking design is that the venture bank pays some premium per dollar of insured loan per year (5% in my modeling), and at closeout, the underwriter receives a large share of the investment equity value, either as stock or cash (50% in my modeling). I do not allow for bankers to purchase multiples of the loan value, nor multiple EDCSs using the same loan. This would be considered fraud.

1 It is true that the asymptotic classical banking multiplier (1/R where R = reserve fraction) is functionally extinct because of the invention of reserve banking. (e.g. Federal Reserve Bank, European Central Bank, etc.). A modern bank can make any loan it thinks is valid, and settle up on the back end. However, the venture-banking proposal is designed to function without requiring access to a central bank except for settlement purposes. The classical multiplier is still in use by regulators to determine reserves, it is just that the Central Banks have replaced gold.

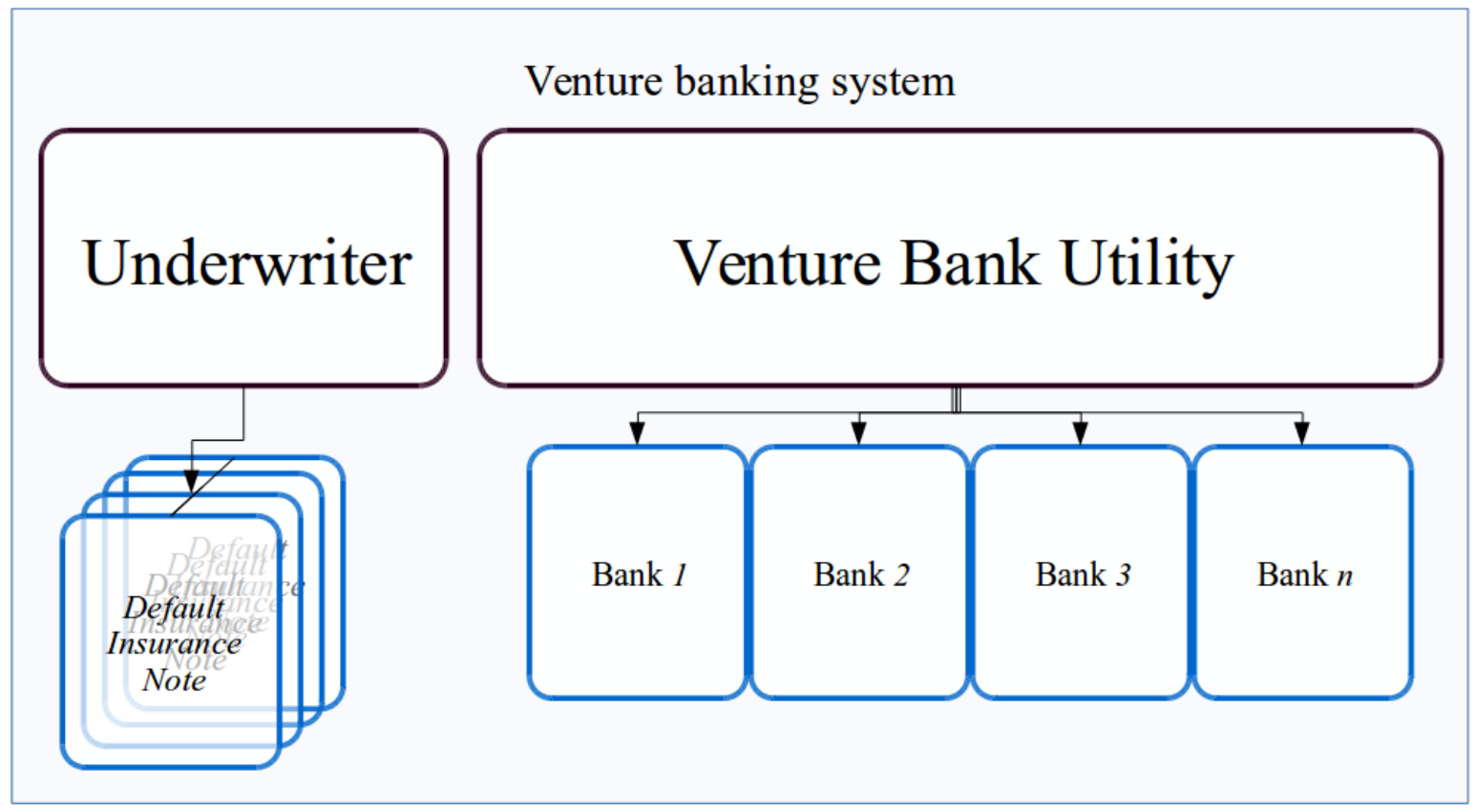

*Figure 2: Proposed venture capital banking system. Venture bank utility (VBU) is paired with an underwriter issuing equity default clawback swap (EDCSs).The clients of the VBU are venture capital firms.*

In the venture-bank system an individual venture capital fund would put their money into a venture bank utility (VBU), which would take care of bank operations for the fund (Fig 2).

## 3 The jackpot problem in venture banking

Figures 3 and 4 show the respective profitability of the venture-bank versus the simple Equity Default Swap (EDS) underwriter in stark terms. The optimum strategy for the venture-bank in the region of normal returns is to ensure that all the investments fail.

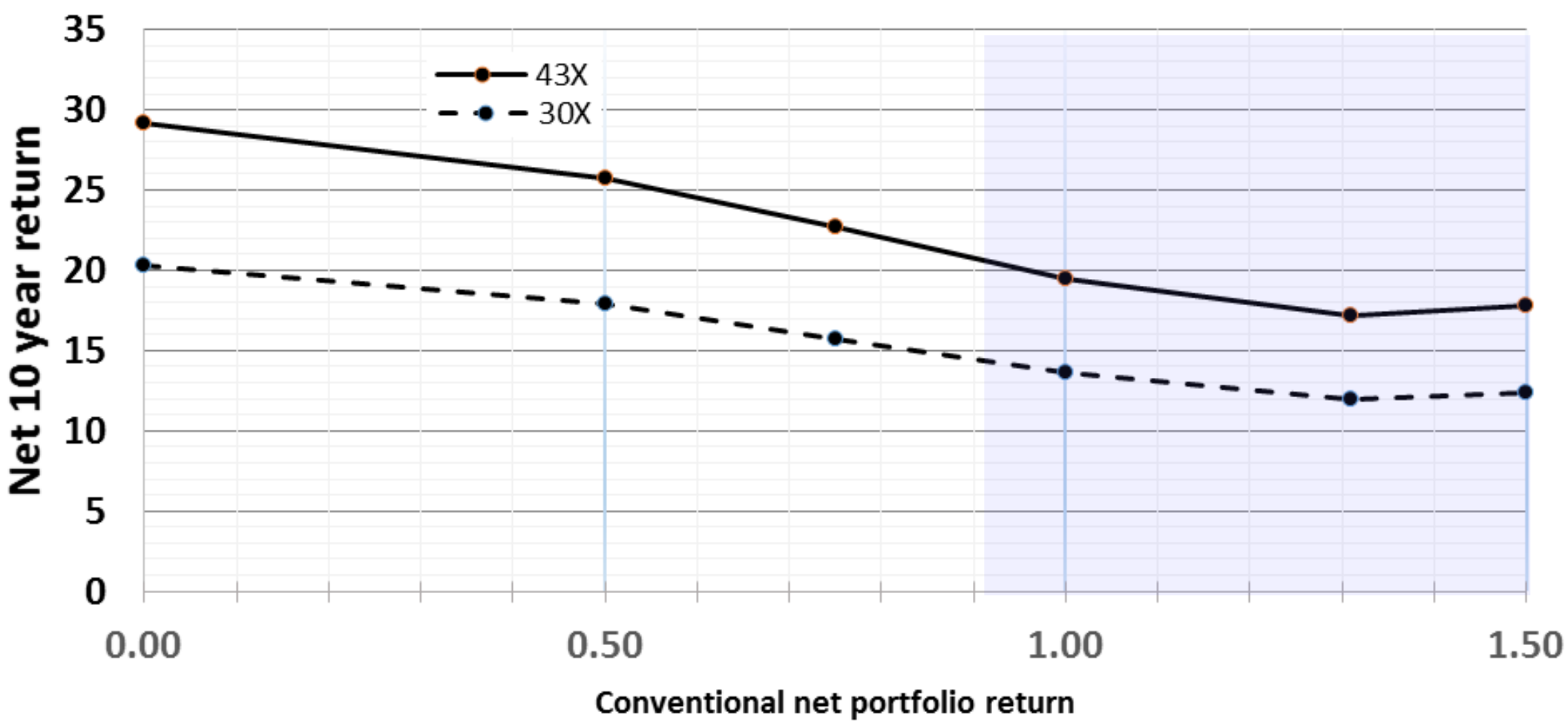

*Figure 3: 5% EDS premium and 50% equity for 100% EDS coverage. Shows net 10 year total return varying what the return of a conventional portfolio would have been without the clawback part of the instrument. Shading indicates normal range of returns. Break-even at 1.0 on the Y axis.*

In figure 3, we see what the return is for a venture-bank without a clawback for an EDS modeled on the CDS contract. Obviously, an underwriter that had 90% or more of venture-banks intentionally causing all their investments to fail would crash like AIG did in 2008.

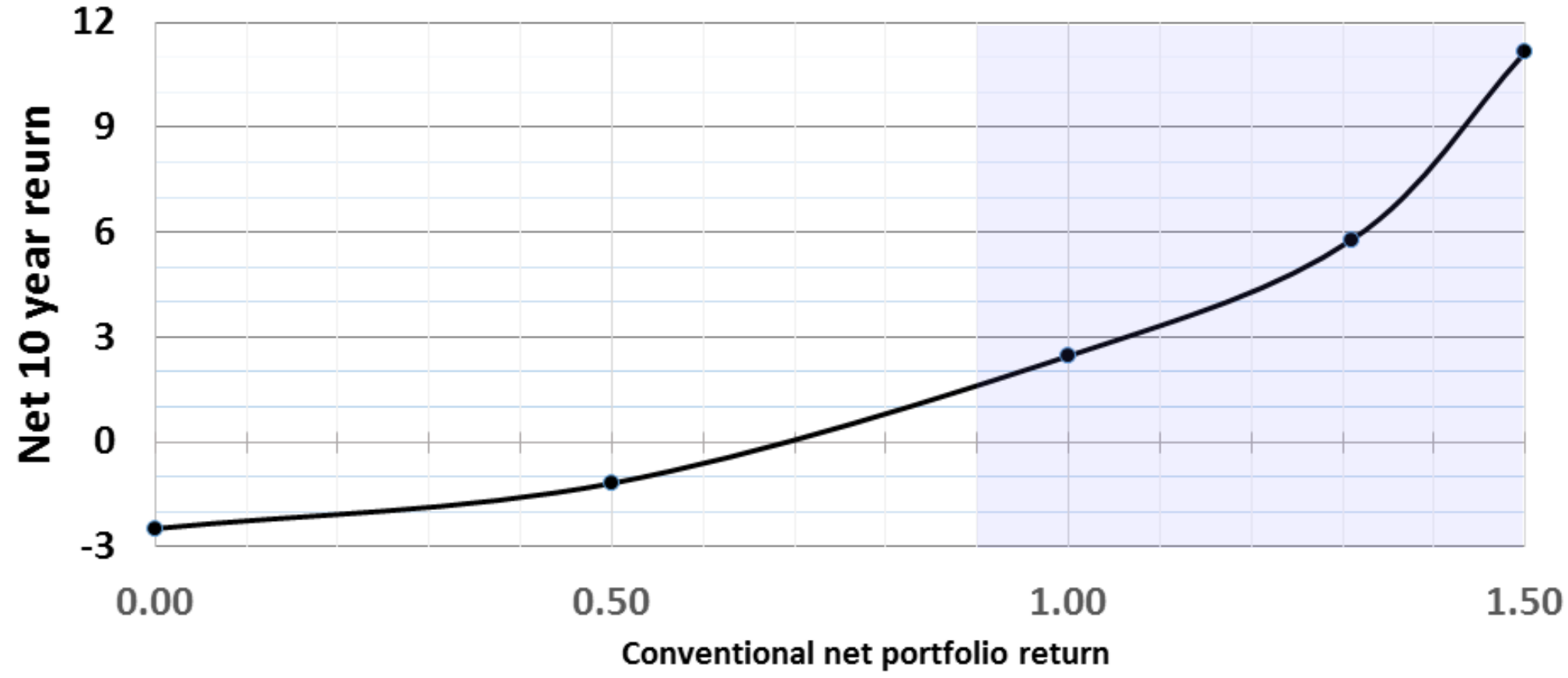

*Figure 4: EDS underwriter's return without clawback. Break-even = 0 on the Y axis, exclusive of operating overhead. Shading indicates normal range of returns. However, this should change because of the perverse incentive.*

Figure 4 shows the underwriter's return with no clawback The returns for EDS underwriters increase dramatically for a conventional VC portfolio return between 1.50 and 2.27, and some individual venture funds would be in that region. The Kauffman data of figure 1 shows returns up to 8.0. This is impossible to graph as return on investment (ROI) beyond a total conventional return of approximately 2.27, because between 2.27 and 2.28, the underwriters no longer have any invested funds. So percentages and fractional returns become meaningless. Instead, it is only possible to graph the earnings per dollar of sold contracts as seen in figure 5. The shape of this figure 5 curve is different from figure 4 because it is earnings per dollar of sold contracts, not return on investment.

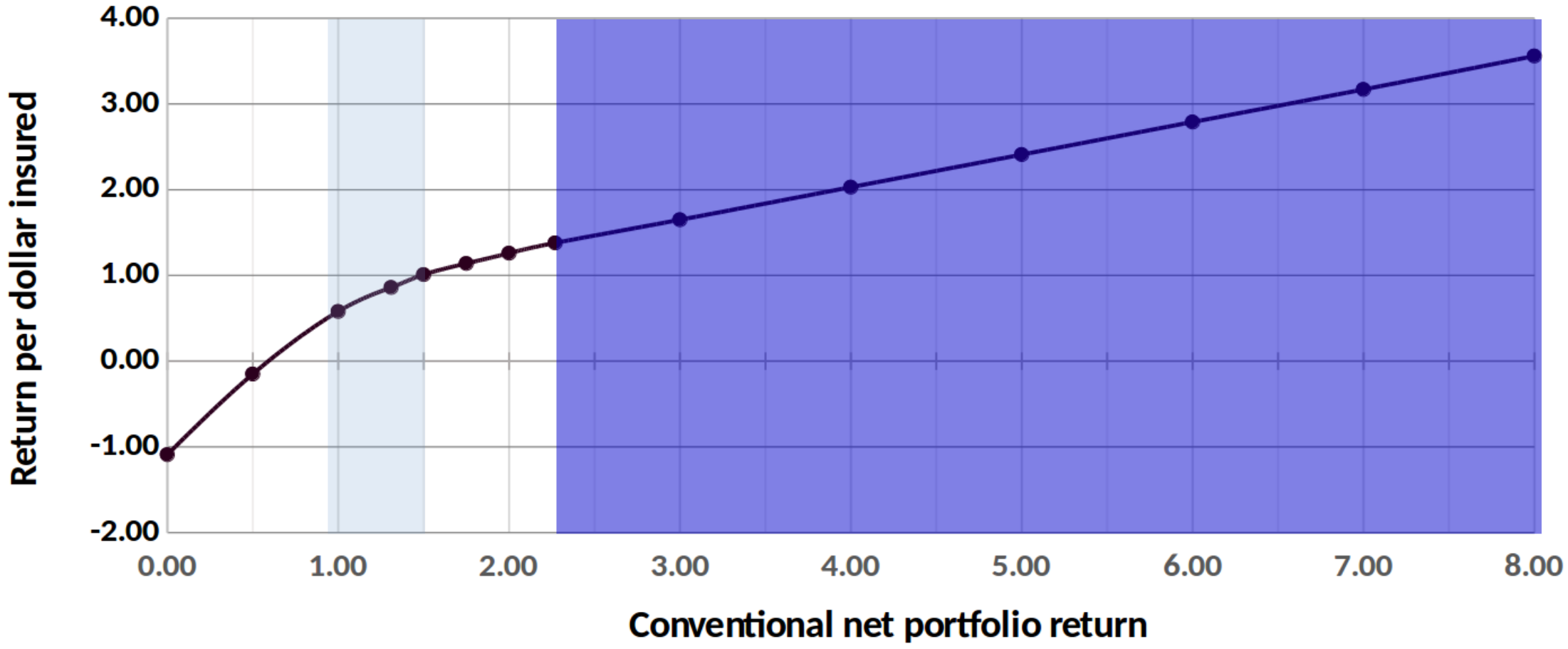

*Figure 5: EDS underwriter return per dollar of sold contracts varying conventional return. Light shading from 0.9 to 1.5 is the same shaded region as shown in figures 3 and 4. Dark shading from 2.27 to 8.00 is the region where underwriters have zero invested funds, and no carrying costs.*

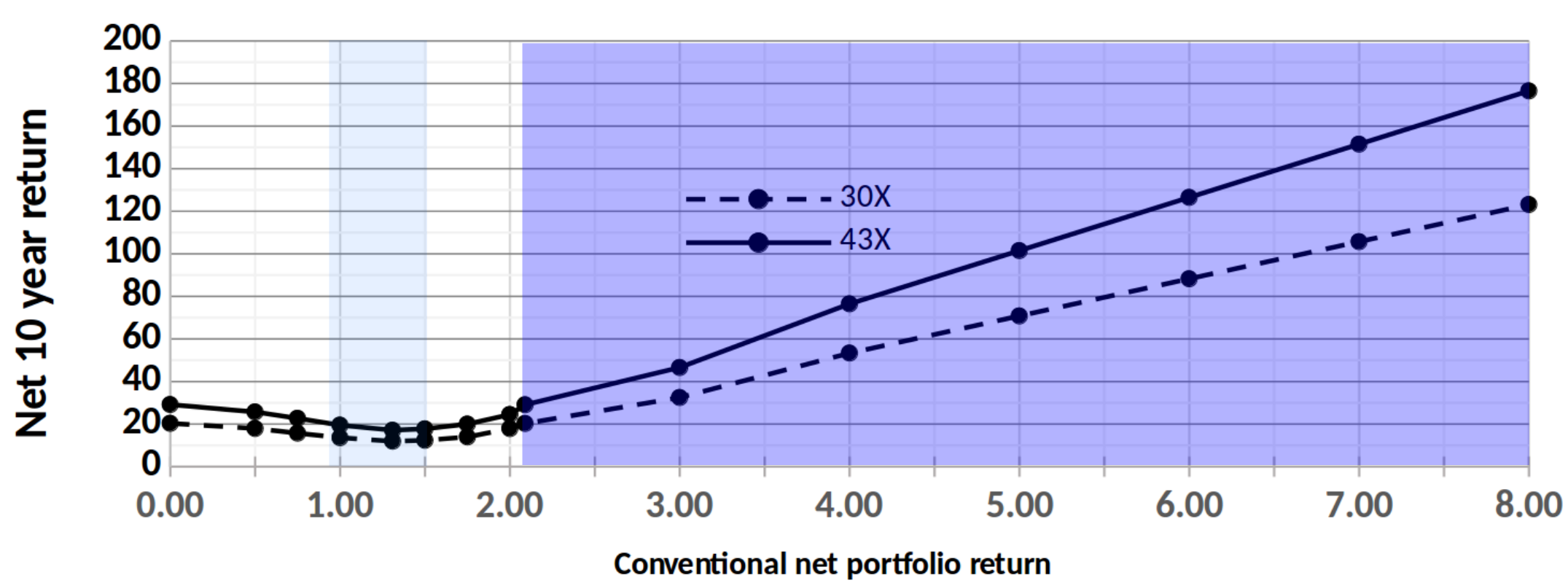

*Figure 6: No clawback. 10 year venture bank return for individual venture capital funds insured with an EDS. Light shading is the overall portfolio return region for the bank as a whole seen in previous figures. Dark shading starts where a venture capital fund's returns pass the zero point.*

As we see in figure 6, without a clawback, the returns for venture-banks can rise above the 29X maximum that an insured would receive from taking all their investments to zero, but this only happens in the highest earnings areas. As seen in figure 1, 90% of VC funds make less than the 2.2 level of return, so it is unrealistic to expect good behavior from VC firms faced with this choice.

For the bank as a whole, and underwriting as a whole, it is unlikely that total average returns above 1.50 would be seen. However, there are venture capital firms that have outstanding returns as reported by Kauffman data. Venture capitalists are gamblers to a significant extent. Playing devil's advocate for the venture capital firms, it might be argued that by using the pure EDS the bottom 90% of venture funds could be eliminated by wise underwriters. In theory, this could result over time in higher returns for everyone, but that would probably occur at the expense of radically shrinking the pool of available capital for entrepreneurs, which is an undesirable result. In addition, it is very unlikely that underwriters would be any better at figuring out which VC firms would be successful than current limited partners can, and they would probably go bankrupt if they tried. So this is not a realistic idea.

I propose this novel instrument of the clawback lien instead to cure both problems.

## 4 Clawback lien merged into EDS instruments creates EDCS

I propose a novel instrument to augment with the simple EDS. This new instrument would create a lien on the ownership of the venture capital firm that would have a term period of months or years. It would be attached to a venture firm's portfolio or else a timeline decided upon when the venture firm should have closed out successful investments. However, the underwriter's lien on the firm could be removed by either payment of the lien's monetary value, or an acceptable transfer of some other asset to the underwriter.

The lien would charge interest at a rate determined by the underwriter's terms on the EDS. The venture capital firm would normally have an option to negotiate with the underwriter for alternative payments such as equity share in another investment in lieu of monetary payment. Merging the clawback lien into the EDS yields an Equity Default Clawback Swap (EDCS).

The clawback lien could vary depending on the structure of the EDCS payments, whether flat, front-loaded or back-loaded, as these result in some variation for the underwriter.

To set the clawback lien rate will require modeling of these features of the EDCS rather than a simple mathematical formula. Based on the most current modeling, the optimum clawback is on the order of 77%. However, other rates were modeled as shown below. With higher clawback rates, underwriters make significantly better returns.

Below is shown the results from a model using a EDCS rate of 5% per year, a EDCS equity share of 50% and a clawback lien rate of 62.3%.

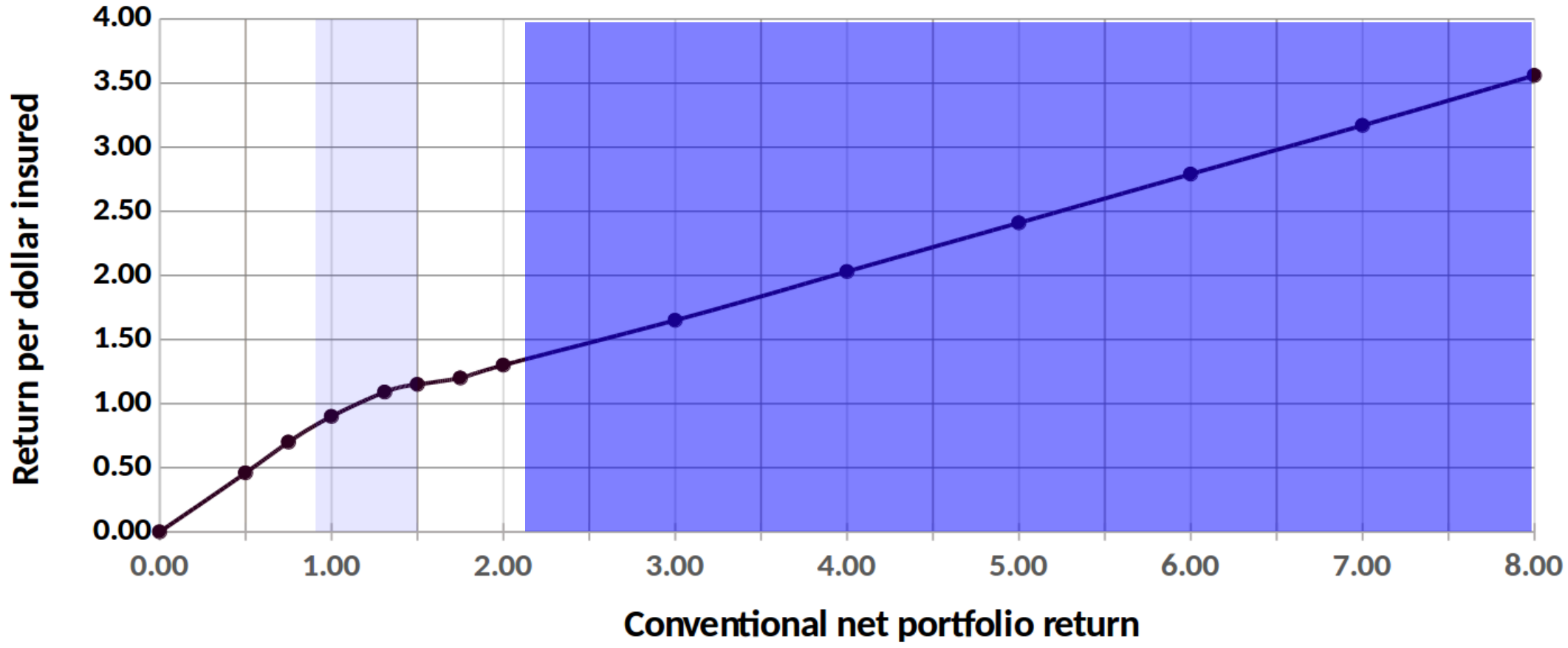

*Figure 7: Clawback lien results for underwriter.* ***Break-even = 0.*** *EDCS rate 5% per year, EDCS equity share 50%, clawback lien rate 62.3%. EDCS underwriter return per dollar invested varying conventional return. Light shading from 0.9 to 1.5 is the estimated normal range of conventional portfolio returns for a venture-bank. Dark shading from 2.27 to 8.00 is the region where underwriters have zero invested funds, and no carrying costs.*

Note that in figure 7, this is not return on investment (ROI). ROI was not modeled here, but it will be higher than figure 4's returns. The shaded region of figure 4 should be used as a guide.

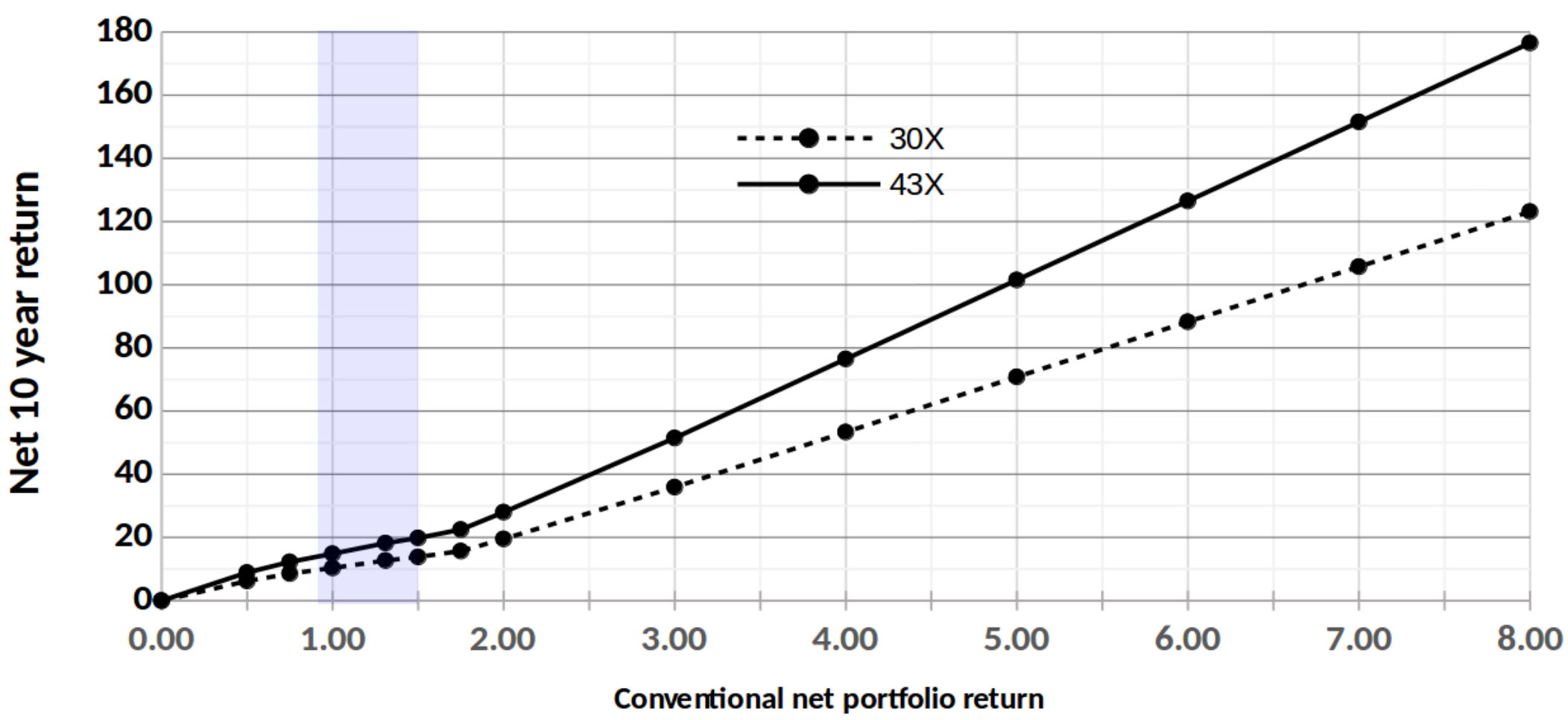

*Figure 8: Clawback lien results for underwriter.* ***Break-even = 1.0****. EDCS rate 5% per year, EDCS equity share 50%, clawback lien rate 62.3%. 10 year venture bank return for individual venture capital funds. Light shading from 0.9 to 1.5 is the estimated normal range of total portfolio returns for a venture-bank.*

As we see in figures 7 and 8, by means of the clawback lien, the perverse incentive is eliminated. The venture capital firm that is using the VBU to make investments now has a smoothly increasing profit curve. The EDCS underwriter would no longer lose money except at the very bottom end.

The goal of clawback lien rate modeling is that when a portfolio for a venture capital firm goes to zero, the EDCS returns are barely negative or do not go below zero. In practice, underwriters will experience losses in such situations because the venture capital firms will simply not have enough assets to cover. But underwriter losses should be minimal nevertheless because of the VBU system, which will ensure that assets are available for defunct firms. As derivative based instruments, EDCS clawback liens should be in primary position.

When clawback liens are in force, I do not expect that venture capital firms would abandon their investments very often when they have major losses. This is both because there is a correlation between past performance and future results, and that those operating the fund will want to maintain good relations with underwriters, and with their VBU.

## 5 Concluding remarks

The clawback lien instrument should all but eliminate losses for underwriters that are insuring venture-bank investments when venture capital firms have returns from zero to break-even. This clawback lien should encourage underwriters providing EDCS coverage to venture capital firms that have experienced significant past losses, except in the worst cases. Venture capital firms composed of new principals should also have minimal difficulty obtaining EDCS coverage with the clawback lien in place. The clawback lien thus creates robust support for venture capital firms, and hence for entrepreneurial ventures, while keeping most losses where they belong – with the venture capitalists.

## 6 Acknowledgements

I want to thank Geoffrey Gardiner for critique on banking and insurance.

## 7 Glossary

AIG – American International Group. A global insurance company providing insurance products to commercial, institutional and individual customers. They also provide mortgage insurance and credit default swap (CDS) contracts.

BACPA – Bankruptcy Abuse Prevention and Consumer Protection Act of 2005. For these purposes, BACPA strengthened the rights of derivative holders to collect immediately.

Basel accords – There are three sets of banking regulations set by the Basel Committee on Bank Supervision. These are known as Basel I, Basel II and Basel III.

CDS – Credit Default Swap. The purchaser makes premium payments to the underwriter and the contract insures a loan on some asset, typically a real estate loan. If the borrower defaults on the loan, then the purchaser is paid the face value of the contract, and transfers the asset to the underwriter.

EDCS – Equity Default Clawback Swap. A proposed derivative that provides for payment on loans that pay back less than face value that are made by venture capitalists as investments, defined in this paper. This EDCS contract is they key to enabling this new type of banking to function.

EDS – Equity Default Swap. The simple form of a proposed derivative that provides for payment on loans that pay back less than face value that are made by venture capitalists as investments, as defined in this paper. This form does not include a clawback, and thus it replicates the weakness seen in the AIG debacle that triggered the 2008 banking crisis.

MOC – Multiple of Original Capital. Some amount of money is put into the bank that is its capital. This amount is enlarged by the Basel accords rules into the complete Tier 1 and Tier 2 capital that is used by the bank as reserves. The total outstanding investments divided by the original capital placed in bank Tier 1 reserves is the MOC.

VBU – Venture Bank Utility. This is a proposed new entity that handles the banking operations for a set of venture banks. See figure 2.